\documentclass{iau}
\usepackage{graphicx}
\usepackage{amsmath}
\usepackage{subcaption}
\usepackage{xfrac}
\title[Non-Gaussian inference] 
{Non-Gaussian inference from non-linear and non-Poisson biased distributed data}

\author[Metin Ata et al.]   
{Metin Ata, Francisco-Shu Kitaura \& Volker M\"uller}

\affiliation{Leibniz Institute for Astrophysics (AIP), \\ An der Sternwarte 16, 14482 Potsdam\\ 
email: {\tt mata@aip.de} \\[\affilskip]}

\pubyear{2014}
\volume{306}  
\pagerange{119--121}
\jname{Statistical Challenges in 21st Century Cosmology}
\editors{ eds.}
\begin{document}

\maketitle

\begin{abstract}
We study the statistical inference of the cosmological dark matter density field from non-Gaussian, non-linear and non-Poisson biased distributed tracers. We have implemented a Bayesian posterior sampling computer-code solving this problem and tested it with mock data based on $N$-body simulations.

\end{abstract}

\firstsection 

\section{Introduction}

The distribution of galaxies poses a challenging multivariate statistical problem.
Galaxies are biased tracers of the underlying three-dimensional dark matter density field.
To accurately infer such fields one needs to account for the non-Gaussian, non-Poisson and non-linear biased distribution of galaxies.
We show that this is possible within the Bayesian formalism by explicitly writing down the posterior distribution including these effects. 
\section{Method}
We rely on the Bayesian framework to express the posterior distribution function (likelihood weighted prior) of matter fields given a set of tracers, a biasing and a structure formation model.
In particular we consider a negative binomial distribution for the likelihood modeling the data (the galaxy field). This permits us to model the over-dispersed galaxy counts (see \cite{Kitaura14} and references therein). The expected galaxy number density is related to the dark matter density through a nonlinear scale-dependent expression extracted from $N$-body simulations (see \cite{cen_and_ostriker}; \cite{dlt}; \cite{Kitaura14}; \cite{Neyrinck14}). In this way we extend the works based on the Poisson and linear bias models (\cite{Kitaura08}; \cite{Kitaura10}; \cite{jk}; \cite{jw}) following the ideas presented in \cite{Kitaura12a} and \cite{Kitaura14}. In particular, we implement these improvements in the \textsc{argo} Hamiltonian-sampling code able to jointly infer density, peculiar velocity fields and power-spectra  (\cite{Kitauraetal12b}).
For the prior distribution describing structure formation of the dark matter field we use the lognormal assumption  (\cite{C_J91}). We note however, that this prior can be substituted by another one, e.~g.~based on Lagrangian perturbation theory (see \cite{Kitaura13}; \cite{Kitauraetal12a}; \cite{jw}; \cite{Wang}; \cite{hess}). Alternatively, one can extend the lognormal assumption in an Edgeworth expansion to include higher order correlation functions (\cite{Colombi}; \cite{Kitaura12b}).
We calculate the posterior distribution on a grid of $N_C$ cells dividing our observed volume. 
The lognormal distribution for the dark matter field $\delta_{\rm M}$ can be written  as a function $\mathcal{P}(\delta_{\rm M}|\mathbf{S}(\{p_C \}))$, 
with $\{p_C \}$ being a set of cosmological parameters that go into the covariance matrix $\mathbf{S}$.
For the likelihood we consider  the negative binomial distribution (NB), so that the probability of observing $N_i$ galaxies in a voxel given the expectation value $\lambda^i=f_{\bar{N}}w^i(1+\delta_{\rm M}^i)^\alpha\exp\left[-\left(\frac{1+\delta_{\rm M}^i}{\rho_\epsilon}\right)^\epsilon\right]$ is given in the following form: $\mathcal{L}(\mathbf{N}|\mathbf{\lambda},\beta)$, with the parameter $\beta$ modeling the deviation of Poissonity. For $\beta\rightarrow\infty$ or for low expectation values (low $\lambda$) the NB tends towards the Poisson distribution function. We note that deviations from Poissonity to model the galaxy distribution have been considered in previous works (see e.~g.~\cite{S_H}; \cite{Sheth}).


The expected number counts $\lambda^i$ is constructed from the mean number per voxel $\bar{N}$, the completeness $w^i$, and the density $(1+\delta_{\rm M}^i)^\alpha \exp\left[-\left(\frac{1+\delta_{\rm M}^i}{\rho_\epsilon}\right)^\epsilon\right]$ in the voxel. $f_{\bar{N}}$ is the normalisation factor: $f_{\bar{N}} = \sfrac{\bar{N}}{ \langle (1+\delta_{\rm M}^l)^\alpha{\rm e}^{\left[-\left(\frac{1+\delta_{\rm M}^l}{\rho_\epsilon}\right)^\epsilon\right]}} \rangle$, where $\langle ... \rangle$ denotes the ensemble average over the whole volume. The biasing parameters are given by \{$\alpha,\beta,\epsilon,\rho_\epsilon $\}.
Combining these terms to a posterior function gives a full description of the desired probability to infer the dark matter field from the observed galaxy distribution  $P(\delta_M|\mathbf{N},\mathbf{S}(\{p_C\})) \propto \mathcal{P}(\delta_{\rm M}|\mathbf{S}(\{p_C \}))  \times \mathcal{L}(\mathbf{N}|\mathbf{\lambda},\beta) $:
\begin{gather}
\label{posterior}
P(\delta_M|\mathbf{N},S(\{p_C\}))  = \frac{1}{\sqrt{(2\pi)^{N_C} \mathrm{det}(\mathbf{S})}} \prod_{l=1}^{N_C} \frac{1}{1+\delta_M^l}  \\ \nonumber \times \exp{\left(-\frac{1}{2} \sum_{ij}\left[(\ln{(1+\delta_M^i)}-\mu^i) S^{-1}_{ij} (\ln(1+\delta_M^j)-\mu^j)\right]\right)}  \\ \nonumber \times  \prod_{l=1}^{N_C}\left( \frac{f_{\bar{N}}w^l(1+\delta_{\rm M}^l)^\alpha{\rm e}^{\left[-\left(\frac{1+\delta_{\rm M}^l}{\rho_\epsilon}\right)^\epsilon\right]} \Gamma(\beta+N^l)}{N^l!\, \Gamma(\beta)\left(\beta+f_{\bar{N}}w^l(1+\delta_{\rm M}^l)^\alpha{\rm e}^{\left[-\left(\frac{1+\delta_{\rm M}^l}{\rho_\epsilon}\right)^\epsilon\right]}\right)^{N^l}
 \left(1+\frac{f_{\bar{N}}w^l(1+\delta_M^l)^\alpha  {\rm e}^{\left[-\left(\frac{1+\delta_{\rm M}^l}{\rho_\epsilon}\right)^\epsilon\right]}}{\beta}\right)^\beta}\right) \, {.}
\end{gather}
Note that for the field $\Phi=\ln(\varrho)-\langle \varrho \rangle = \ln(1+\delta_{\rm M}) -\mu$ our prior is exactly a Gaussian prior. Therefore, our method yields at the same time the optimal  logarithmic Gaussianised density field (\cite{Neyrinck09,CS}).
\section{Validation}
\begin{figure}
        \centering
        \begin{subfigure}[b]{0.4\textwidth}
                \includegraphics[width=\textwidth]{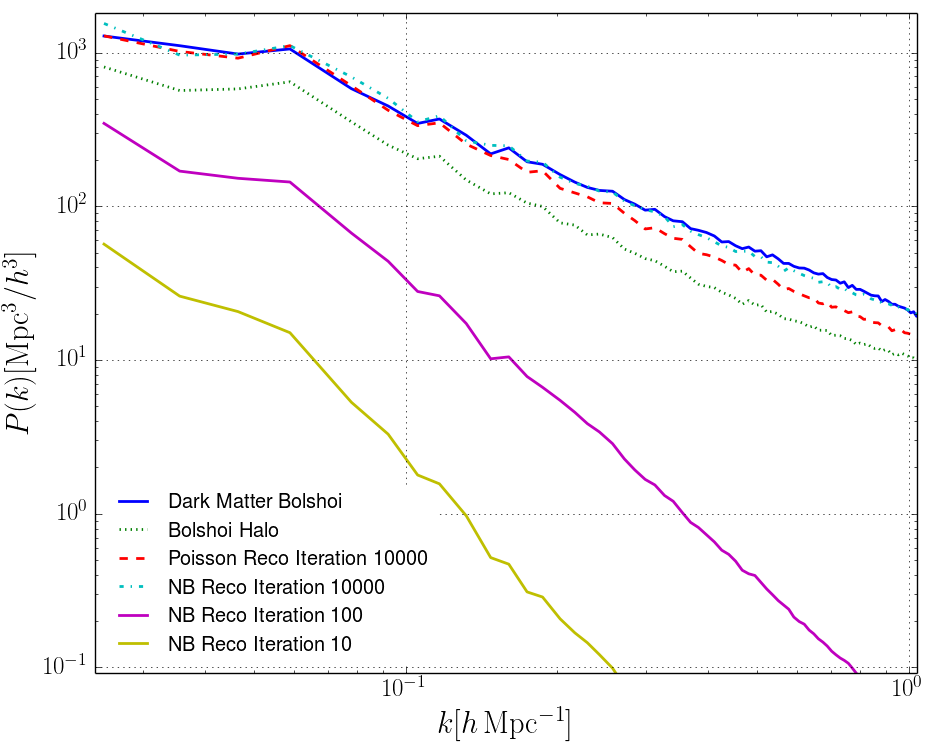}
                \caption{Power spectra}
                \label{fig:power}
        \end{subfigure}%
        ~ 
        \begin{subfigure}[b]{0.4\textwidth}
                \includegraphics[width=\textwidth]{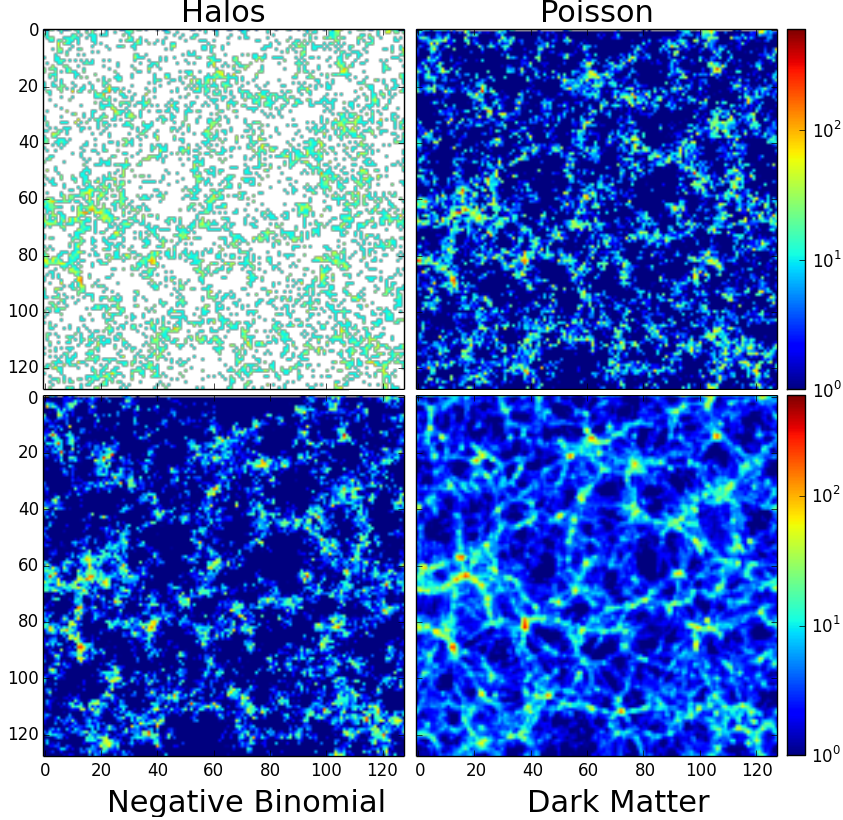}
                \caption{Slices ob the observed volumes}
                \label{fig:slices}
        \end{subfigure}
        ~ 
\end{figure}

We employ our recently developed Hamiltonian Markov Chain Monte Carlo based computer code for Bayesian inference from the posterior distribution shown in Eq. \ref{posterior}  (see Ata et al in prep).
To test our method we take a subsample of $2\cdot 10^5$ halos from the halo catalog of the Bolshoi dark matter only $N$-body simulation at redshift zero comprising the halo masses between $10^9$ and $10^{14} ~\rm M_\odot$ (\cite{klypin}).
We run two independent chains: the first one with the classical Poisson-Lognormal model, and the second one with the novel NB-Lognormal model. Our results demonstrate that our new model is able to recover the dark matter density field yielding unbiased power-spectra (within 5\%) in the $k$-range of 0.02 to 0.6 $h$ Mpc$^{-1}$  (see solid blue and bashed dotted light blue lines in Fig.~\ref{fig:power}). However, the Hamiltonian sampling run with the Poisson-Lognormal model including the same nonlinear deterministic bias produces biased reconstructions with power-spectra deviating  about 20 \% at   $\sim k=0.4$  $h$ Mpc$^{-1}$  (see red dashed line in Fig.~\ref{fig:power}). The superior three-dimensional resamblance between the original dark matter field from the $N$-body ($8\cdot10^9$ dark matter particles) and the reconstruction  (based on $2\cdot 10^5$ halos), using the NB likelihood as compared to the Poisson likelihood, is apparent in the slice cuts shown in Fig.~\ref{fig:slices}.
\section{Conclusions}
We have introduced a detailed posterior distribution within the Bayesian framework accurately modeling the statistical nature of the distribution of galaxies. Moreover, we have implemented a Hamiltonian sampling code to infer the corresponding matter density fields.
We validated our method against realisitic mock data for which the underlying dark matter density field is known. Our numerical tests emphasize the importance of a scale-dependent nonlinear bias and the deviation from Poissonity.

\end{document}